\documentclass[conference]{IEEEtran}
\IEEEoverridecommandlockouts
\usepackage{cite}
\usepackage{amsmath,amssymb,amsfonts}
\usepackage{algorithmic}
\usepackage{multicol}
\usepackage{graphicx}
\usepackage{textcomp}
\usepackage{xcolor}
\usepackage{float}
\usepackage{subfigure}
\usepackage{caption}
\usepackage{geometry}
\usepackage[font=small,labelfont=bf]{caption}
\def\BibTeX{{\rm B\kern-.05em{\sc i\kern-.025em b}\kern-.08em
    T\kern-.1667em\lower.7ex\hbox{E}\kern-.125emX}}
\begin{document}
\newcommand{\norm}[1]{\left\lVert#1\right\rVert}
\vspace{5pt}
\title{High-Precision Machine-Learning Based Indoor Localization with Massive MIMO System}
\vspace{-5pt}
\author{Guoda~Tian, 
        Ilayda Yaman,
        Michiel Sandra,
        Xuesong Cai, 
        Liang Liu, 
        Fredrik Tufvesson
        \\Dept. of Electrical and Information Technology, Lund University, Sweden  \\
        email: firstname.lastname@eit.lth.se \\

        \thanks{
        
        This work has been supported by Ericsson AB, Horizon Europe Framework Programme under the Marie Skłodowska-Curie grant agreement No.,101059091, the Swedish Research Council (Grant No. 2022-04691), and the strategic research area ELLIIT, Excellence Center at Linköping — Lund in Information Technology. The authors would like to thank Lund University Humanities Lab, and the LTH robotic lab for the help with equipment. Special thanks to Henrik Garde, Alexander Dürr, Steffen Malkowsky, Sara Willhammar and Sirvan Abdollah Poor, Nikhil Challa, and Ove Edfors for their assistance throughout the measurements. 
        
        }
}

\maketitle
\begin{abstract}
	
High-precision cellular-based localization is one of the key technologies for next-generation communication systems. In this paper, we investigate the potential of applying machine learning (ML) to a massive multiple-input multiple-output (MIMO) system to enhance localization accuracy. We analyze a new ML-based localization pipeline that has two parallel fully connected neural networks (FCNN). The first FCNN takes the instantaneous spatial covariance matrix to capture angular information, while the second FCNN takes the channel impulse responses to capture delay information. We fuse the estimated coordinates of these two FCNNs for further accuracy improvement. 
To test the localization algorithm, we performed an indoor measurement campaign with a massive MIMO testbed at 3.7\,GHz.  
In the measured scenario, the proposed pipeline can achieve centimeter-level accuracy by combining delay and angular information. 

\end{abstract}

\begin{IEEEkeywords}
    massive MIMO, machine learning, radio-based localization, measurement, pre-processing 
\end{IEEEkeywords}\vspace{-3pt}

\section{Introduction}\vspace{-3pt}
{C}{ellular}-based localization opens the door for numerous location-aware applications such as navigation, autonomous driving, robot mapping, intelligent transportation \cite{b1,b2,b3,b4,b5}, etc. Since cellular-based localization has a crucial function in those applications, efforts have been made by standardization organizations, such as the third generation partnership project (3GPP) to further investigate this technology. 
Motivated by the growing demand for high-quality localization services, cellular-based localization technology has rapidly evolved in the last ten years. In the current 3GPP specification (Release 17 and 18), cellular-based localization is one of the fundamental features of the new radio standard.

Traditional cellular-based localization methods include time-of-arrival (ToA), time-of-flight (ToF), time-difference-of-arrival (TDoA), angle-of-arrival (AoA) based methods \cite{b7}. These methods estimate the user equipment (UE) coordinates directly based on relevant propagation channel characteristics. 
To further enhance localization precision, advanced radio technologies such as ultra-wideband (UWB) \cite{b8} or massive multiple-input multiple-output (MIMO) \cite{b9,b10,b11} have been considered. A UWB system can inherently capture the propagation channels with high delay resolution. 
Similarly, by scaling up the number of antennas at the base station (BS) side, one can achieve a higher angular resolution and thus better AoA estimation quality. Therefore, massive MIMO enables excellent localization accuracy, even though available bandwidth is limited \cite{b9}. Despite the good positioning accuracy delivered by traditional localization algorithms, it is still challenging to use those advanced signal processing techniques because of the computational complexity of the algorithms \cite{b7}. In addition, AoA estimation algorithms usually require array calibration. Since radio-based localization is in itself a regression task (from channel state information to UE positions), data-driven algorithms such as machine learning (ML) can also be utilized \cite{Debast1, Debast2, b12,GP2,Joao,TVT1,Globecom1, IEEEAccess, Bayesian}.  The performance of ML-based localization tasks hinges to large extent on two factors: selecting appropriate channel fingerprints as training data and choosing suitable learning algorithms. Training data can be either the raw transfer function itself \cite{Debast1, Debast2}, or other channel fingerprints such as received signal strength, power delay profile (PDP), angle-delay spectrum, etc, \cite{b12,GP2,Joao,TVT1,Globecom1}. Several ML-based localization algorithms can be found in literature, which are categorized into the Kernel family \cite{GP2}, the neural network family \cite{Debast1, Debast2,b12,Joao,TVT1,Globecom1,IEEEAccess} and the Bayesian family \cite{Bayesian}.  

Applying ML to massive MIMO systems for localization purposes is a promising yet immature research topic since it is challenging to process the rich spatial channel information captured by a massive antenna array.  
Early research work \cite{Joao} proposed an approach by first extracting the delay-angle spectrum as a channel fingerprint and feeding this fingerprint to a convolutional neural network (CNN). This work was validated by simulation. 
The work in \cite{IEEEAccess} proposed an auto-encoder-based calibration algorithm to first calibrate the antenna array before estimating the AoA. Measurement results show that this approach requires fewer training samples. 

In this paper, we analyze a new ML-based localization algorithm for massive MIMO systems and we evaluate our algorithm using real measurement data. The first step of the algorithm is to extract both the instantaneous spatial covariance matrix and channel impulse responses (CIRs) of all antennas as fingerprints. These fingerprints contain channel information in both the angle and the delay domains. Calibration of all RF chains is not necessary for the generation of those fingerprints. We then feed the fingerprints to two separate processing branches consisting of fully connected neural networks (FCNN), to estimate independent UE coordinates. The measurements indicate that those two processing branches deliver similar localization performance and their positioning errors have a weak correlation, given that we have enough training samples. Therefore, we can achieve an accuracy gain by averaging those two coordinates compared to only using the raw transfer function itself. In addition, the measurement results illustrate that pre-processing is crucial to improve localization accuracy. 

\section{System Model and Problem Formulation}
\label{SM}
We consider a single-input multiple-output (SIMO) orthogonal frequency-division multiplexing (OFDM) system, where signals are transmitted by a single-antenna UE to an $M$-antenna BS. At the BS side, each antenna is connected to an individual radio frequency (RF) chain so that the signals from antenna can be processed coherently. Furthermore, we consider an indoor scenario where the UE moves at a normal walking speed, and the $2$-D coordinates of UE at position $\mathbf{p}_i \in \mathbb{R}^2$ are denoted as $\mathbf{p}_i(x_i, y_i)$. 
We can then write the received channel transfer function matrix $\mathbf{Y}_{p_i}\in \mathbb{C}^{M\times N}$ with respect to all subcarriers at position $\mathbf{p}_i$ as, \vspace{-3pt}
\begin{equation}
    \mathbf{Y}_{p_i} = \mathbf{H}_{p_i} \odot\mathbf{\Gamma} + \mathbf{N},
    \label{M1} \vspace{-2pt}
\end{equation}
where the propagation channel between UE and BS at position $\mathbf{p}_i$ is denoted as $\mathbf{H}_{p_i} \in \mathbb{C}^{M \times N}$. $\mathbf{\Gamma} \in \mathbb{C}^{M \times N}$ collects all coefficients which contain amplitudes and phase drifts with respect to all $M$ RF chains and $N$ sub-carriers, while $\mathbf{N} \in \mathbb{C}^{M \times N}$ illustrates an additive noise matrix. In addition, $\odot$ represents the Hadamard product. While the UE moves, we capture in total $\mathcal{T}$ different channel snapshots. Thus,  $\mathcal{T}$ different channel matrices $\mathbf{Y}_{p_i}$ and their corresponding UE coordinates $\mathbf{p}_i(x_i,y_i)$ are recorded. Our target is to establish a functional relationship between $\mathbf{Y}_{p_i}$ and coordinates $\mathbf{p}_i$. 

\section{Proposed Localization method}
\label{PLm}
First, we calculate instantaneous spatial covariance matrices as well as truncated CIRs as two separate fingerprints to reflect angular properties and delay properties, respectively.  We then train two neural networks, which learn functional relationships between those two extracted fingerprints and their corresponding $2$-D UE coordinates. As a final step, the two estimated UE coordinates are averaged to obtain the final estimated UE positions. \vspace{-3pt}   
\subsection{Generate Fingerprint}
The performance of ML-based localization algorithms can be significantly enhanced if  adequate channel fingerprints are extracted from the raw received channel matrices \eqref{M1}. In accordance with \eqref{M1}, it is a non-trivial task to directly attain calibrated fingerprints such as angle spectrum because of the influence of $\mathbf{\Gamma}$. Considering this, we first compute the instantaneous spatial correlation matrices $\mathbf{C}_{Y,p_i} \in \mathbb{C}^{M\times M}$, by correlating across all subcarriers as   
\begin{equation}
    \mathbf{C}_{Y,p_i} = \mathbf{Y}_{p_i}\hspace{1pt}\mathbf{Y}_{p_i}^H. \\
  \label{M2}  
\end{equation}
Note that $\mathbf{C}_{Y,p_i}$ %
is a Hermitian matrix according to its definition. Therefore, the information embedded in the upper minor diagonal elements is identical to the information of the lower minor diagonal elements. To reduce algorithm complexity, we define a matrix $\hat{\mathbf{C}}_{Y, p_i}\in \mathbb{R}^{M \times M}$ and 
a vector $\hat{\mathbf{c}}_{Y,p_i}\in \mathbb{R}^{M^2}$ as \vspace{-2pt}
\begin{eqnarray}
\begin{aligned}
    \hat{\mathbf{C}}_{Y, p_i} &= \textrm{ltril}\bigl\{\Re\big[{\mathbf{C}_{Y,p_i}}\big]\bigl\} + \textrm{sltril}\bigl\{\Im\big[{\mathbf{C}_{Y,p_i}}\big]\bigl\} \\
    \hat{\mathbf{c}}_{Y, p_i} &= \textrm{vec}\bigl\{\hat{\mathbf{C}}_{Y,p_i}\bigl\},\\
\end{aligned}
\label{reducecomplexity}
\end{eqnarray}
where $\Re\{.\}$ and $\Im\{.\}$ represent real and imaginary parts of a matrix. In addition, $\textrm{ltril}\{.\}$ denotes a matrix operation that replaces all the values above the diagonal as zero and keeps the rest elements, while $\textrm{sltril}\{.\}$ maintains all elements below the diagonal and replaces all other elements (including the diagonal) as zero, and the $\textrm{vec}\{.\}$ operation vectorizes a matrix as a vector. 

Channel information in the delay domain is not presented in $\hat{\mathbf{c}}_{Y, p_i}$ while this information is still useful for enhancing positioning accuracy. Therefore, we generate a CIR matrix $\mathbf{D}\in \mathbb{C}^{M \times L}$ by performing inverse discrete Fourier transform along each row of $\mathbf{Y}_{p_i}$ and then selecting the first $L$ delay bins. We then define a vector $\mathbf{d}\in \mathbb{R}^{2ML}$ that contains all elements of $\mathbf{D}$. Specifically, $\mathbf{d}=[\textrm{vec}\{\Re(\mathbf{D})\}^T,\textrm{vec}\{\Im(\mathbf{D})\}^T]^T$.  Both vectors $\mathbf{d}$ and $\hat{\mathbf{c}}_{Y,p_i}$ are utilized by our proposed ML-based localization pipeline to estimate UE positions.  

\subsection{Proposed processing pipeline}
\begin{figure}[h]
	\centering  
    \includegraphics[width=0.85\linewidth]{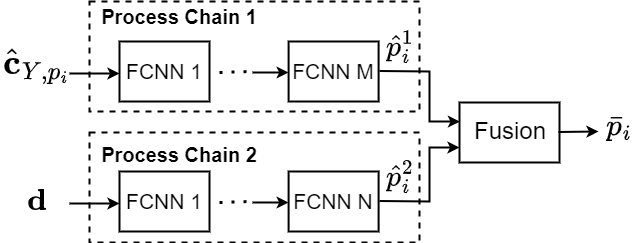}\\
	\caption{Proposed localization pipeline.}
	\label{NNstruc}\vspace{-5pt}
\end{figure}

We use a supervised learning pipeline based on pure FCNN for indoor localization. The framework of our pipeline is illustrated in Fig.
\ref{NNstruc}. \vspace{-5pt}
\begin{figure*}[h]
	\centering  
  \includegraphics[width=1.0\textwidth,height = 4.5cm]{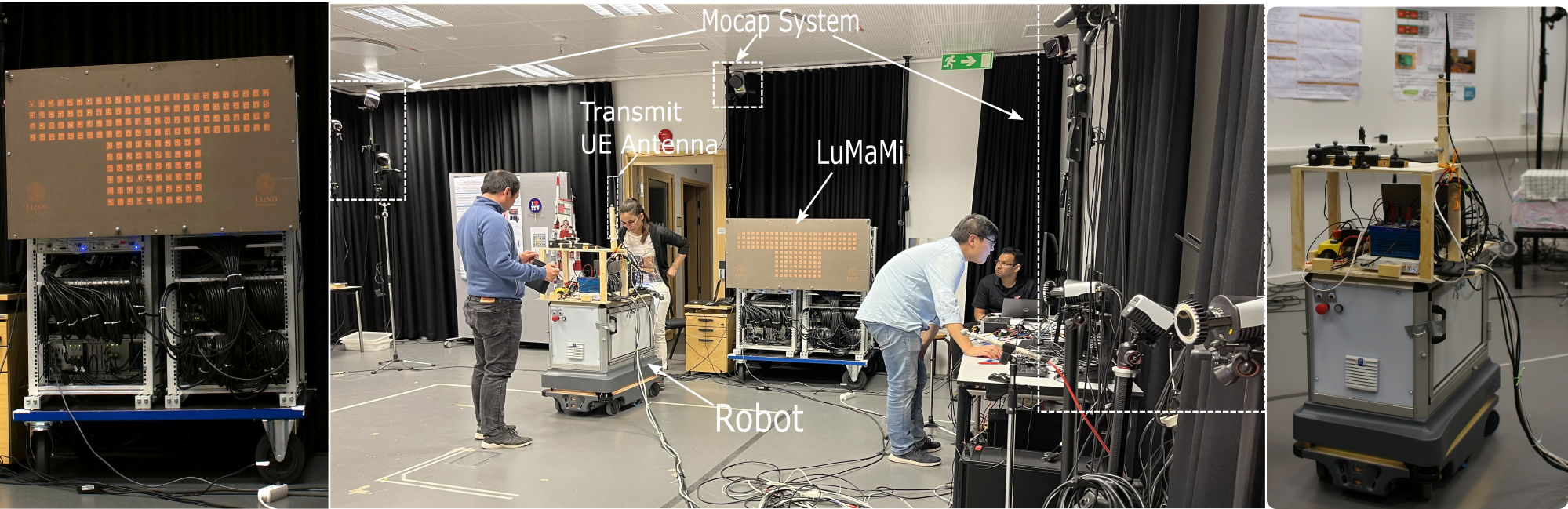}\\
\caption*{(a) BS LuMaMi \hspace{100pt} (b) Measurement Scenario  \hspace{125pt} (c) UE on the robot}
	\caption{Components in the indoor measurement campaign.}\vspace{-10pt}  
	\label{Lab}
\end{figure*}

The pipeline has two separate sub-chains and each chain individually learns the functional relationship between its own input channel fingerprint and UE coordinates. During the training phase, the upper chain takes fingerprint $\hat{\mathbf{c}}_{Y,p_i}$ as input and generates $\hat{\mathbf{p}}_i^1\in \mathbb{R}^2$ as the estimated UE coordinates. In addition, the lower processing chain utilizes fingerprint $\mathbf{d}$ and  generates instead $\hat{\mathbf{p}}_i^2\in \mathbb{R}^2$ as estimated UE coordinates.  We define $\mathbf{f}_{1}(.)$ and $\mathbf{f}_{2}(.)$ as the functions to represent the upper and lower FCNN-based processing chain, respectively, while their individual hyperparameters are contained in vectors $\boldsymbol{\theta}_{1}$ and $\boldsymbol{\theta}_{2}$. In addition, all available training datasets are represented by $\Omega$ while the UE groundtruth label is denoted as a $2$-D vector $\mathbf{p}_i$. Based on this information, we can express two loss functions $\Psi_{1}$ and $\Psi_{2}$, which measure the mean square errors (MSE) between the predicted labels $\hat{\mathbf{p}}_i^1$, $\hat{\mathbf{p}}_i^2$ and groundtruth $\mathbf{p}_i$, as 
\begin{eqnarray}\vspace{-2pt}
     \Psi_{1} = &\sum_{\hat{\mathbf{c}}_{Y,p_i}\in \Omega}||\hspace{1pt}\mathbf{f}_{1}\hspace{1pt}(\hat{\mathbf{c}}_{Y,p_i}, \boldsymbol{\theta}_{1})- \mathbf{p}_i\hspace{1pt}||_F^2, \\
     \Psi_{2} = & \hspace{-31pt}\sum_{\mathbf{d}\in \Omega}||\hspace{1pt}\mathbf{f}_{2}\hspace{1pt}(\mathbf{d}, \boldsymbol{\theta}_{2})- \mathbf{p}_i\hspace{1pt}||_F^2
     \label{Los1},
\end{eqnarray}
where $||.||_F$ denotes the Frobenius norm of a vector, which is used to measure the Euclidean distance between estimated positions and groundtruth labels. The hyperparameter vectors $\boldsymbol{\theta}_{1}$ and $\boldsymbol{\theta}_{2}$ are adjusted through a backpropagation procedure to minimize $\Psi_{1}$ and $\Psi_{2}$ during the training phase. Due to page limitations, we refrain from showing the full-length mathematical derivations, however, readers can find relevant materials in \cite{Simon}. Next, we fuse two individual estimates $\hat{\mathbf{p}}_i^1$ and $\hat{\mathbf{p}}_i^2$. Considering the algorithm complexity, we combine $\hat{\mathbf{p}}_i^1$ and $\hat{\mathbf{p}}_i^2$by taking their average as the final estimated UE coordinate $\bar{\mathbf{p}}_i$, i.e, $\bar{\mathbf{p}}_i = (\hat{\mathbf{p}}_i^1+\hat{\mathbf{p}}_i^2)/2$. We show through the measurement results that an apparent localization accuracy enhancement can readily be achieved by this combination.

\section{Massive MIMO Measurement Campaign}
\label{Measure}
To verify the localization pipeline, an indoor measurement campaign was conducted inside Lund University Humanities Lab's motion capture (MoCap) studio.  During this measurement campaign, the Lund University massive MIMO testbed (LuMaMi) \cite{LUMAMI} acted as a BS while a robot carried the UE equipped with a dipole antenna. While the robot was moving, the 3-D coordinates of the UE antenna were tracked by the MoCap system continually with millimeter accuracy. The environment in MoCap studio, LuMaMi,  and the robot are shown in Fig. \ref{Lab}. We briefly introduce the measurement devices and scenarios, for a detailed description of the measurement campaign please see \cite{luvira}.
\subsection{Measurement devices}
\vspace{-3pt}
During our measurement campaign, the UE moved along a pre-defined trajectory and the BS was equipped with $100$ individual RF chains. To extract more information from the \textit{azimuth} compared to the \textit{elevation} domain, the upper $4\times 25$ antenna elements are selected and each antenna is connected to an individual RF chain. Those antenna elements were separated by approximately half wavelength ($4$\,cm) in both the vertical and horizontal directions. Antenna indexes are labeled as follows: antennas on the first row are labeled in turn from 1 to 25 (from left to right), the second row from 26 to 50, the third row from 51 to 75 and the last row from 76
to 100. 
Both UE and BS operated at a carrier frequency of 3.7\,GHz and occupied 20\,MHz bandwidth. An OFDM tranmission scheme was used, and $N = 100$ sub-carriers were allocated for uplink pilots. The UE  transmitted pilot signals every $10$ ms, which were then  received by the BS to estimate the channel transfer function in \eqref{M1}. 
While the UE was moving, the coordinates of the UE antenna were continually computed by the MoCap system, which consisted of $18$ networked high-speed infrared (IR) cameras. The MoCap system updated the coordinates of the interesting object (UE antenna) at a frame rate of $100$ Hz, which was the same as the channel snapshot rate.  
\begin{figure}[htbp]
  \centering
  \includegraphics[width=0.7\linewidth]{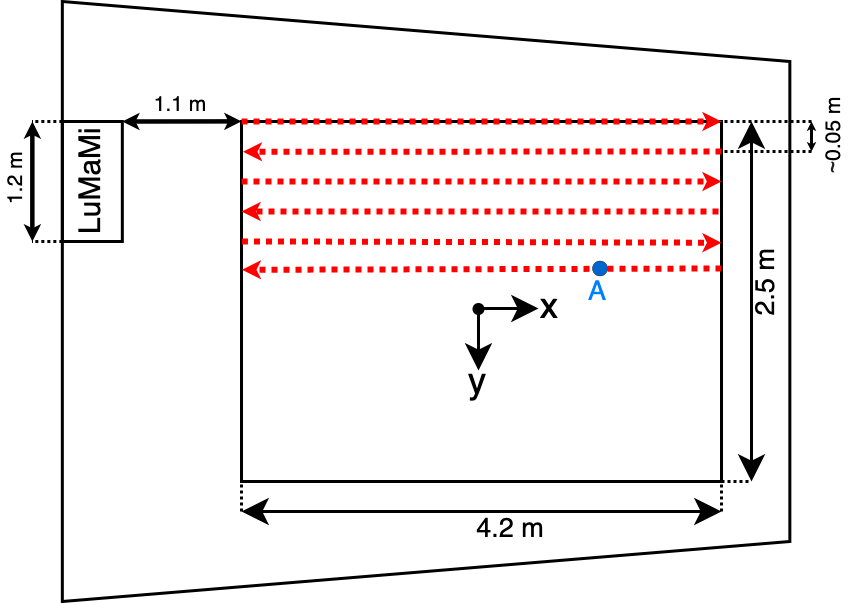}
  \caption{Measurement arrangement in the MoCap studio and the selected point A (0.216, -1.335) in the effective area which is by 4.2\,m $\times$ 2.5\,m. The red dotted arrow shows the trajectory of the UE.}
  \label{Diagram}\vspace{-7pt}
\end{figure}

\subsection{Measurement scenarios and Procedures}
The measurement route consisted of 75 parallel UE  trajectories, as illustrated in Fig. \ref{Diagram}, which resulted in 75 individual channel measurements. For the $i$-th trajectory,  $\mathcal{T}_i$ channel snapshots were recorded. We denote a complex tensor $\mathcal{A}_i \in \mathbb{C}^{\mathcal{T}_i\times M\times N}$ to store those channel snapshots. The whole measurement campaign collects $\mathcal{T} = \sum_i \mathcal{T}_i = 302500$ channel snapshots. We define a complex tensor $\mathcal{A}' \in \mathbb{C}^{\mathcal{T} \times M \times N}$ that combines all $\mathcal{A}_i$, which represents the overall dataset. 

In the beginning of the measurement campaign, the robot was placed at the border of our measurement area. After finishing each trajectory, the robot was moved around $5$ cm along the y-direction, while its orientation remained unchanged. During each measurement, the robot moved straigh in the x-direction, with a speed of $0.1$ m/s. The same measurement procedure was repeated $75$ times so that the whole measurement area was densely scanned with a resolution of around $5$ cm in the y-direction and $1$ mm in the x-direction.  Throughout all measurements,  all objects (except the robot) inside the room remained static and humans were not present in the direct vicinity. \vspace{-2pt} 

\section{Results and Discussions}
\label{RD}
\subsection{Measured channel response}
\begin{figure}
 \begin{subfigure}
  \centering
  \includegraphics[width=0.87\linewidth ]{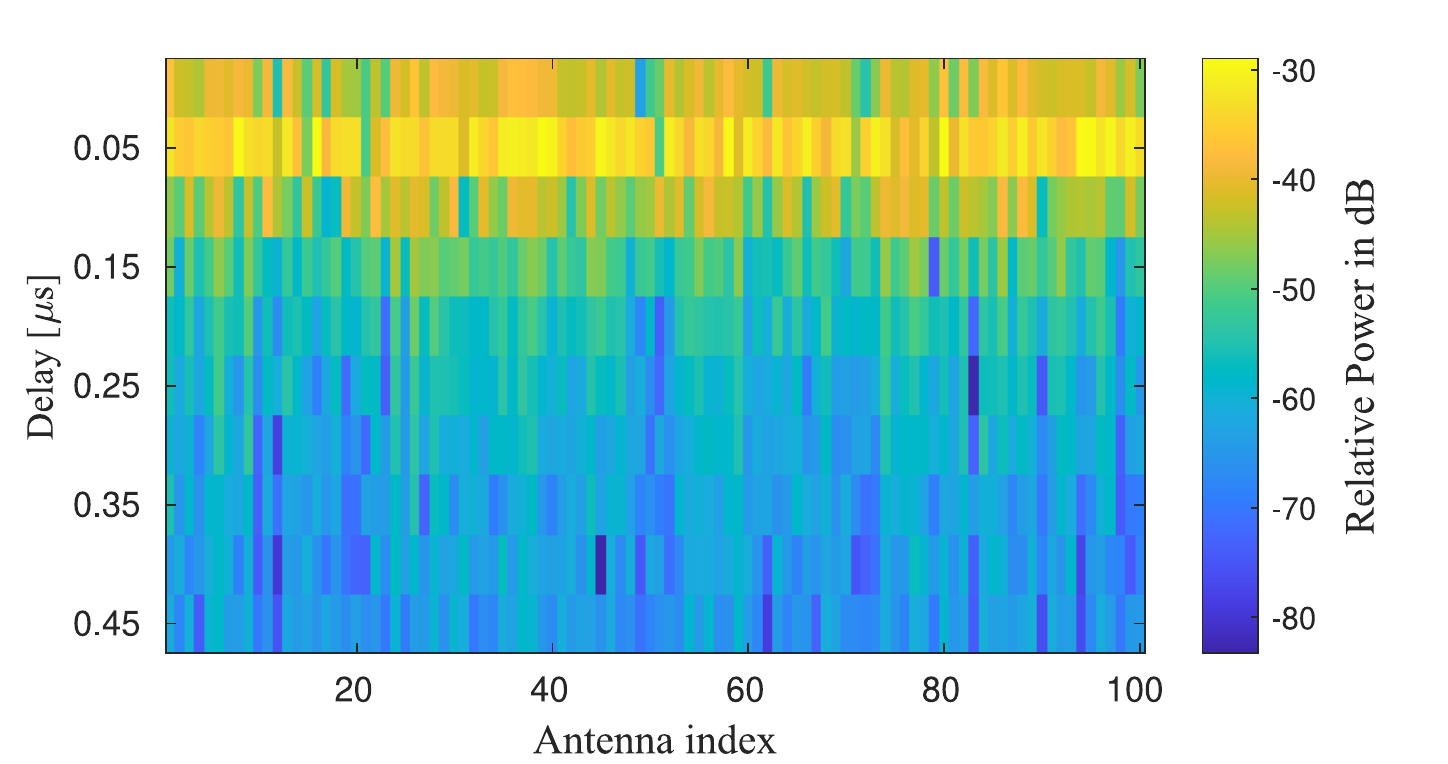}\vspace{-5pt}
\caption*{\hspace{-25pt}(a) Power delay profile}
 \end{subfigure}
 \begin{subfigure}
  \centering
  \includegraphics[width=0.87\linewidth]{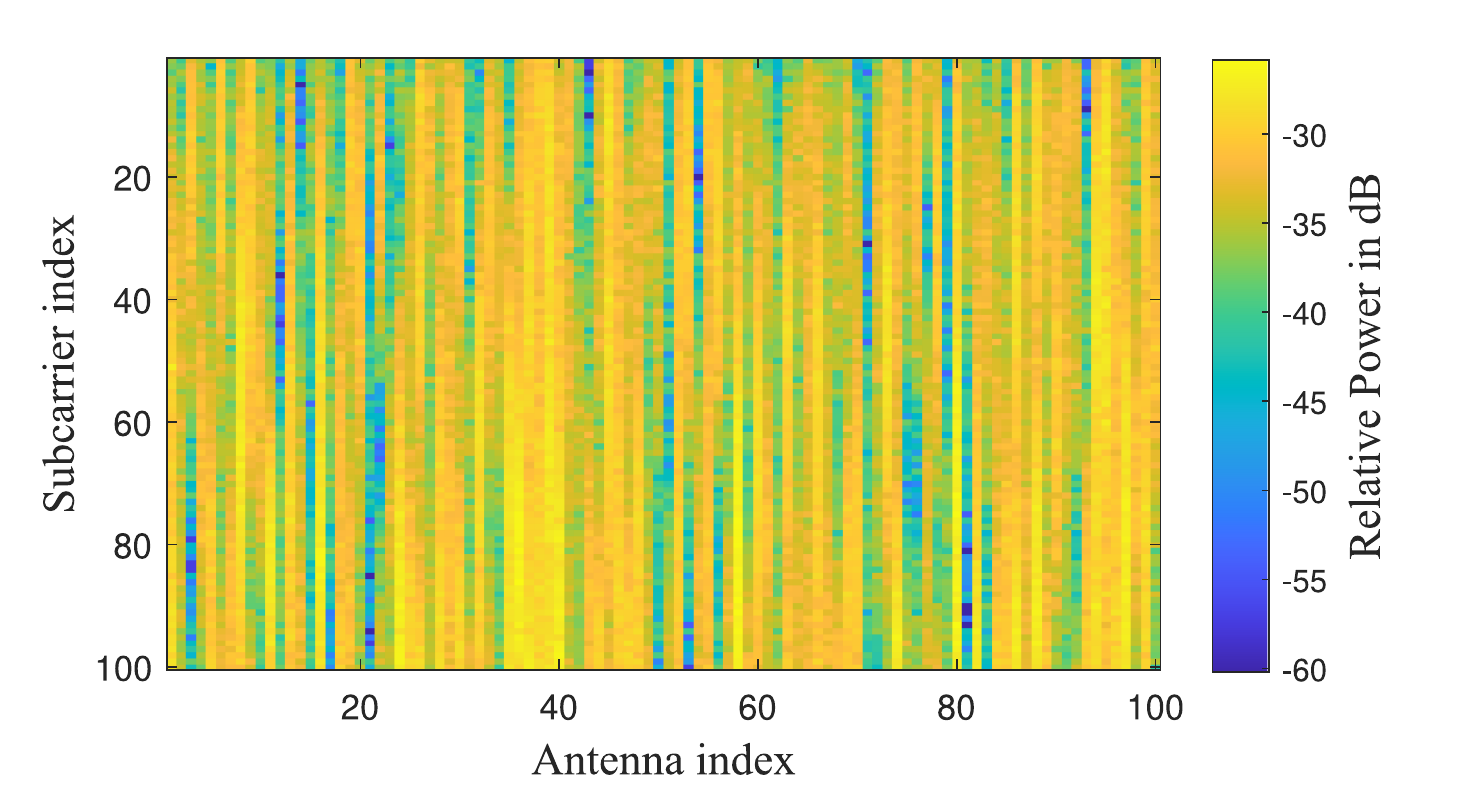}\vspace{-5pt}
  \caption*{\hspace{-25pt}(b) Power of the transfer function}
 \end{subfigure}
   \caption{Power delay profile and power of the transfer function at position A.}\vspace{-5pt}
\label{channel}   
\end{figure}

\vspace{-3pt}
To visualize the measured indoor massive MIMO channel characteristics, one UE position is selected and labeled in Fig.~\ref{Diagram} as position A. We illustrate some key channel characteristics at this position, such as the PDP and the power over the different subcarriers for all antennas in Fig. \ref{channel}.  This figure shows typical indoor channel characteristics since the majority of power in the delay domain is concentrated in the first few delay bins. The same characteristics are also inherent in the transfer functions, with much higher variations of signal power among different antennas while the power among different subcarriers at every single antenna varies more smoothly and slowly. In this propagation environment, it is expected to have a higher frequency correlation compared to spatial correlation.    

\begin{figure}
 \begin{subfigure}
  \centering
  \includegraphics[width=0.87\linewidth]{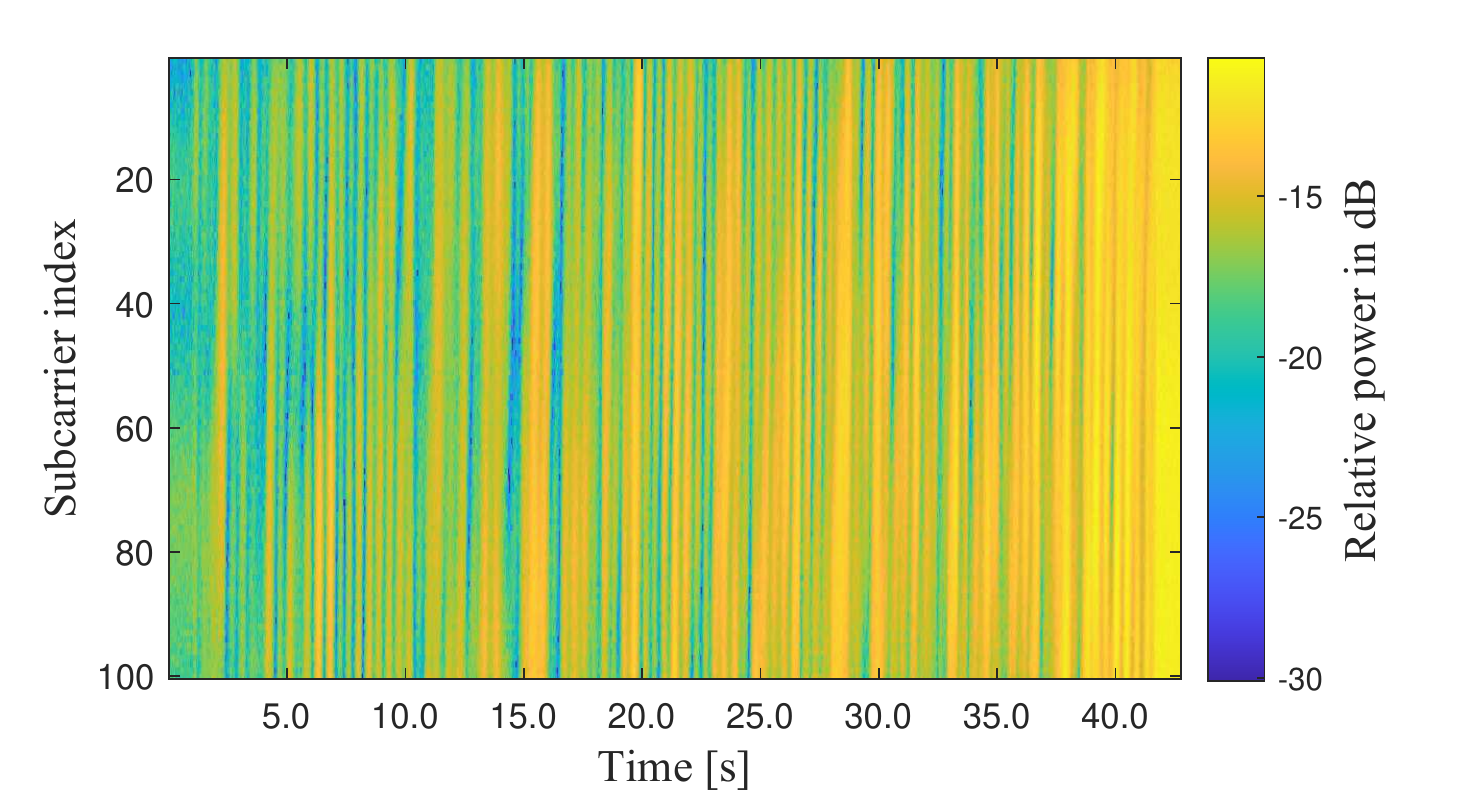}\vspace{-5pt}
\caption*{\hspace{-20pt}(a) Amplitude variation of the first antenna}
 \end{subfigure}
 \begin{subfigure}
  \centering
  \includegraphics[width=0.87\linewidth]{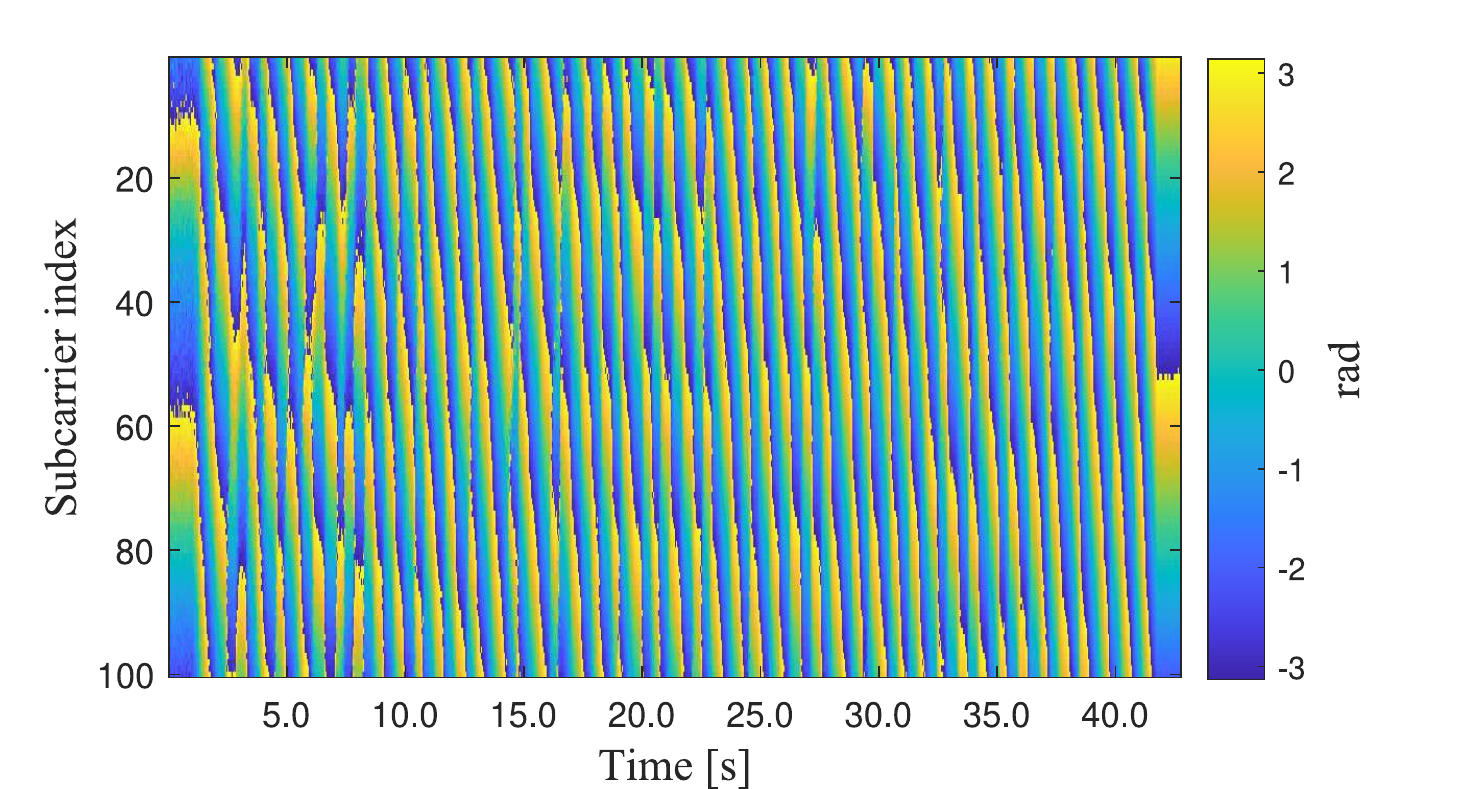}\vspace{-5pt}
  \caption*{\hspace{-20pt}(b) Phase variation of the first antenna}
 \vspace{-3pt}
 \end{subfigure}
   \caption{Amplitude and phase of the transfer function at the first antenna along the UE trajectory.} \vspace{-7pt}
\label{Raw_TF}   
\end{figure}
\begin{figure}[htbp]
\vspace{-5pt}
  \centering
  \includegraphics[width=0.85\linewidth]{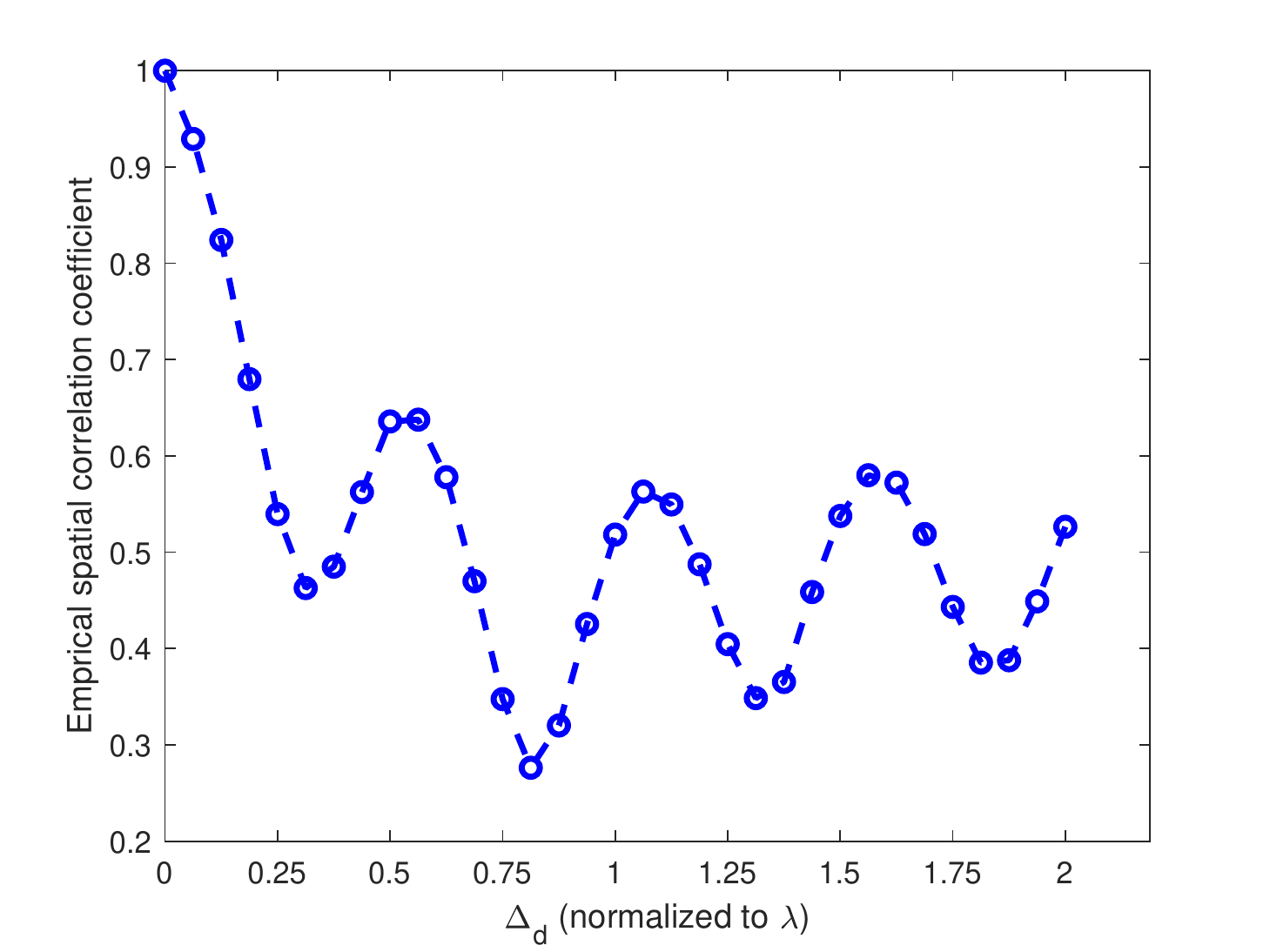} 
  \caption{Empirical spatial correlation function with respect to different UE separation distances. $\lambda\hspace{1pt}\approx\hspace{1pt}8.1$ cm and $\Delta_d$ represents the physical distance, which is normalized to $\lambda$, between two positions.} \vspace{-6pt}
  \label{Empi}
\end{figure}

To display the change of channel response when the UE is moving, we select the first antenna and plot channel transfer functions with respect to all subcarriers along one robot moving trajectory in Fig. \ref{Raw_TF}. As seen from the phase and amplitude, the UE remains static during the first and last second. We can also observe that received signal power increases, which indicates that the UE approaches the BS.  In addition, we investigate the channel spatial correlation for different UE positions. We denote the received channel transfer function at position $p_x$ as $\tilde{\mathbf{H}}_{p_x} \in \mathbb{C}^{M \times N}$ and we define a vector $\tilde{\mathbf{h}}_{p_x} \in \mathbb{C}^{MN}$ by vectorizing $\tilde{\mathbf{H}}_{p_x}$. Suppose that we consider in total $\mathcal{T}'$ positions, the spatial correlation coefficient $\rho\hspace{1pt}(\Delta_d)$, is estimated as  \vspace{-4pt}  
\begin{equation}
  \rho\hspace{1pt}(\Delta_d) = \frac{1}{\mathcal{T}'}\sum_{p_x}\Bigg\{\frac{\tilde{\mathbf{h}}_{p_x}^H\tilde{\mathbf{h}}_{p_{x+\Delta_d}}}{\sqrt{||\tilde{\mathbf{h}}_{p_x}||^2_F ||\tilde{\mathbf{h}}_{p_{x+\Delta_d}}||^2_F}}\Bigg\},   
  \label{Expectation}\vspace{-2pt}  
\end{equation}
where $\Delta_d$ represents the separation distances of two UE positions.
 We select the first UE trajectory and plot the absolute value of $\rho\hspace{1pt}(\Delta_d)$ with respect to $\Delta_d$ from $0$ to $2 \lambda$ according to (\ref{Expectation}) in Fig. \ref{Empi}, where $\lambda$ represents the wavelength. As illustrated in Fig. \ref{Empi}, the spatial correlation is still strong if  $\Delta_d \leq \frac{1}{8}\lambda$. However, a significantly weaker correlation is expected for larger separations. \vspace{-6pt}
\subsection{Preprocessing and Network Parameter setting}

\begin{table}
\footnotesize
\centering
\caption{Parameter settings of the proposed neural network.}\vspace{-10pt}
\begin{center}
\begin{tabular}{p{1.7cm}p{2.0cm}p{1.8cm}p{1.45cm}}
\hline\hline
& Network 1 & Network 2 & Network 3 \\
\hline
Input layer & $2MN\times MN$ & $M^2\times M^2$ & $2ML\times ML$   \\
Hidden layer 1& $MN\times MN/2$ & $M^2\times M^2/2$ & $ML\times ML$\\
Hidden layer 2& $MN/2\times MN/4$ & $M^2/2\times M^2/4$ & $ML\times ML$ \\
Hidden layer 3& $MN/4\times MN/4$ & $M^2/4\times M^2/4$ & $ML\times 512$ \\
Hidden layer 4& $MN/4\times 1024$ & $M^2/4\times 1024$ & $512\times 512$ \\
Hidden layer 5& $1024\times 512$ & $1024\times 512$ & $512\times 256$   \\
Hidden layer 6& $512 \times 128$ & $512 \times 128$ & $256 \times 128$  \\
Hidden layer 7& $128 \times 32$ & $128 \times 32$ & $128 \times 32$  \\
Hidden layer 8& $32 \times 4$ & $32 \times 4$ & $32 \times 4$  \\
Output Layer & $4 \times 2$ & $4 \times 2$ & $4 \times 2$ \\
Network Size         & $1.01$ GB   & $349$ MB & $440$ MB \\
Batch Size   & 64 & 64& 64  \\        
Epoch & 200 & 200 & 200 \\
Complexity & $O(M^2N^2)$ & $O(M^4)$ & $O(M^2L^2)$\\
\hline\hline
\end{tabular}
\end{center}
\label{table: Network Structure}\vspace{-13pt}
\end{table}
After collecting the measurement results as tensor $\mathcal{A}'$, the entire tensor is normalized by multiplying a scaling factor $\alpha\ = \sqrt{\frac{\mathcal{T}MN}{||\mathcal{A}^{'}||^2_F}}$ so that the average power of each element of the new tensor $\tilde{\mathcal{A}}\in \mathbb{C}^{\mathcal{T}\times M \times N}  = \alpha\mathcal{A}'$ equals 1. 
As a next step, the total $\mathcal{T}$ samples are divided into a training set with $\mathcal{X}$ training samples and a testing set with $\mathcal{T}-\mathcal{X}$ testing samples. 

To compare our proposed method with the baseline, in total three networks are trained, corresponding to $3$ different training inputs, namely, (1) the transfer functions $\mathbf{Y}_{p_i}$ itself (baseline), (2) the covariance matrix for the whole array, (3) the channel impulse response at first $L = 10$ delay bins. We illustrated the structures of those three FCNNs in Table  \ref{table: Network Structure}. For all those three FCNNs, Leaky Relu is chosen as a non-linear activation function so that the gradient vanish problem \cite{LRelu} can be avoided. We initially set learning rates for all $3$ networks as $0.0001$ and  they are reduced $20\%$ every $10$ epoch. As shown in Table \ref{table: Network Structure}, compared to Network 1, $65\%$ and $55\%$ less storage resources are required by Network 2 and Network 3, respectively. We also investigate the algorithm complexities of those three networks. Compared to the baseline, a reduction of time complexity can be achieved by either using the covariance matrix or the truncated CIR, because $L \leq N$ and most of the commercial systems fulfill $M \leq N$. Furthermore, if we input both the covariance matrix and the truncated CIR, the time complexity is $O(M^4+M^2L^2) = O(M^2 \max(M^2, L^2))$, which is still better than feeding the raw transfer function as the input. \vspace{-5pt}

\subsection{Localization Result}\vspace{-2pt}
\begin{figure}[t]
  \centering
  \includegraphics[width=0.85\linewidth]{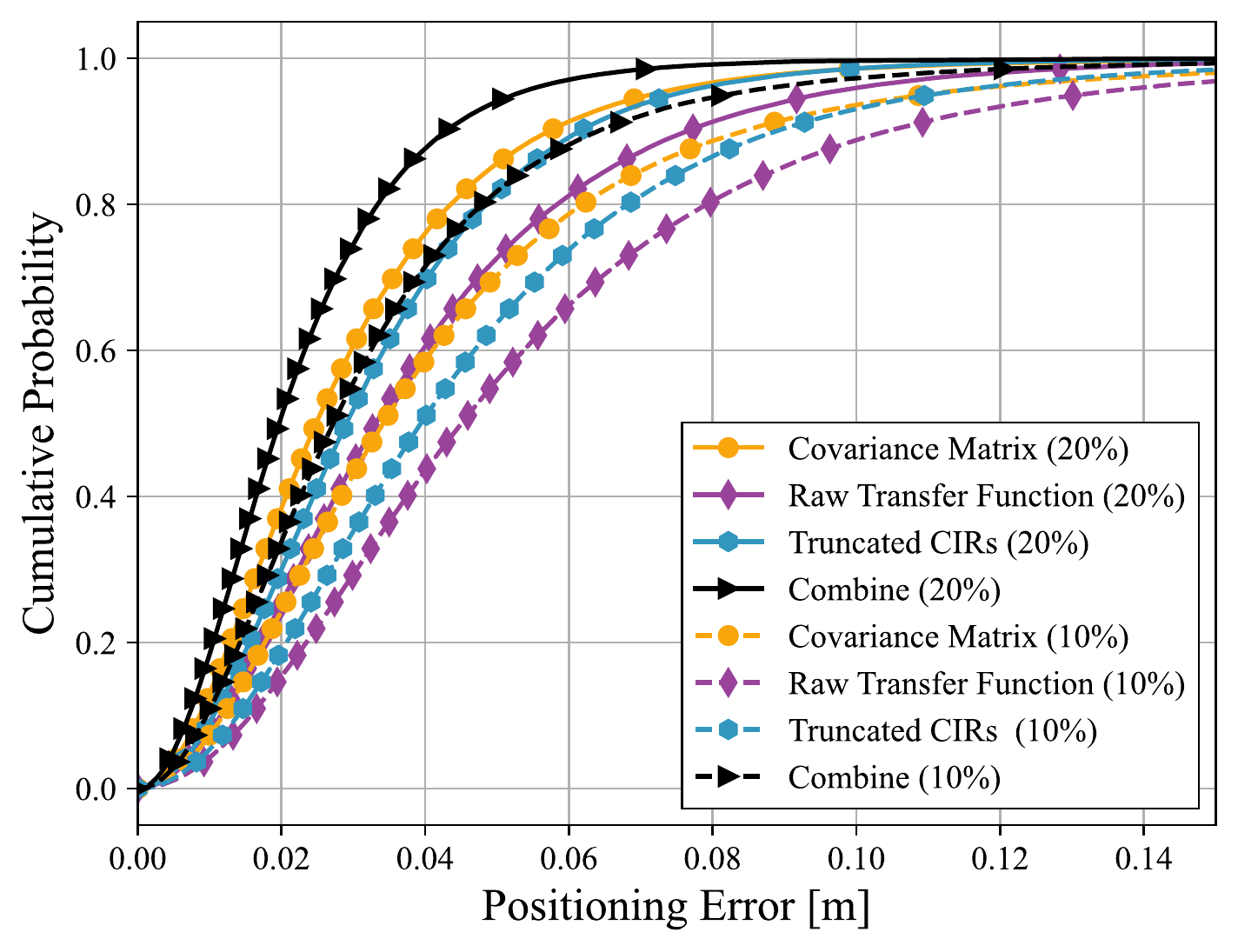} 
  \caption{Positioning error cumulative distribution function with 10\% and 20\% training data, respectively. } \vspace{-10pt}
  \label{Pos1}
\end{figure}
\begin{figure}
  \centering
  \includegraphics[width=0.85\linewidth]{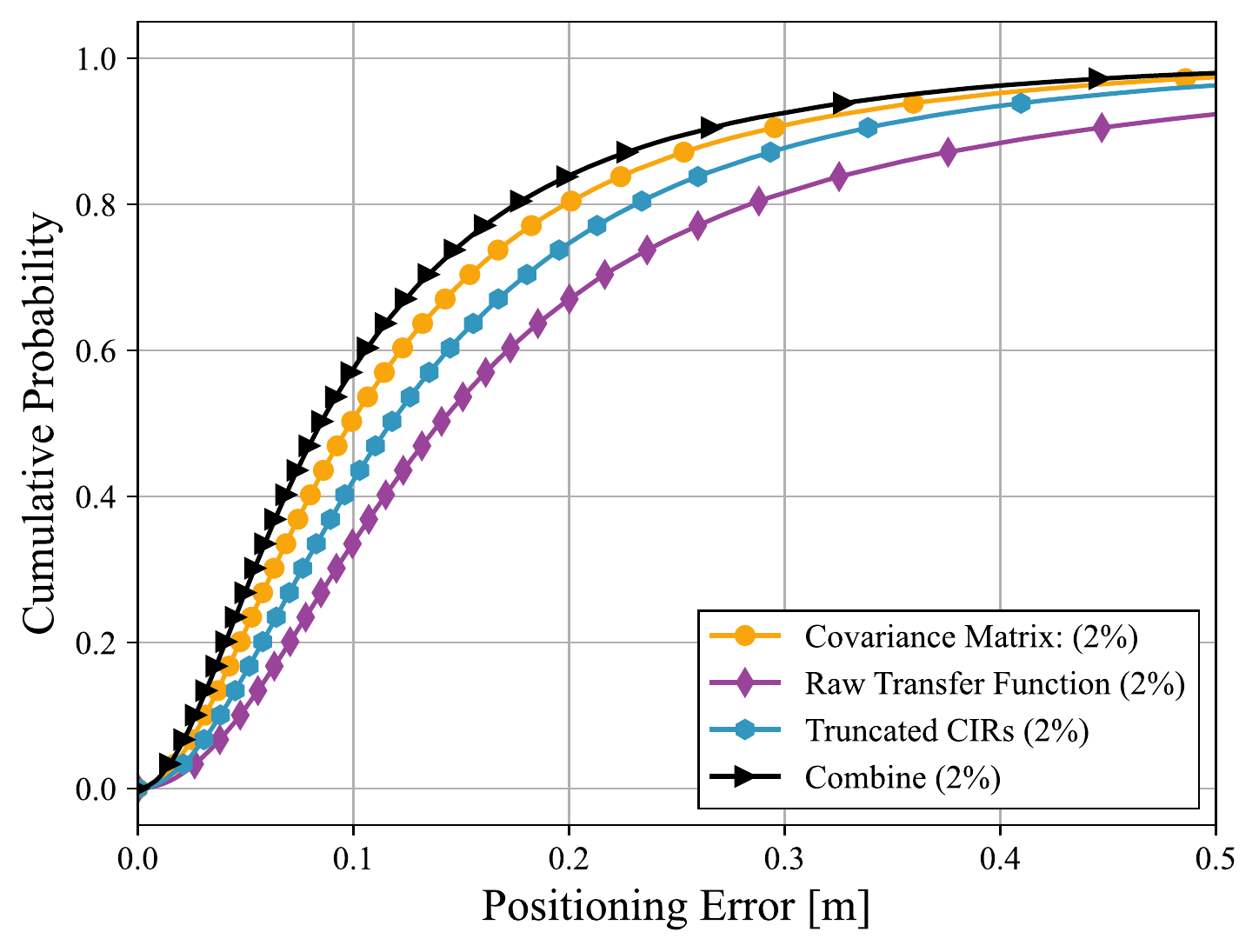}\vspace{-4pt} 
  \caption{Positioning error cumulative distribution function with 2\% training data. }\vspace{-15pt}
  \label{Pos2} 
\end{figure}
We finally compare localization results of four localization methods, including \textit{i)} training on the instantaneous covariance matrix, \textit{ii)} training on the truncated CIRs, \textit{iii)} training on the raw transfer functions, \textit{iv)} average combining the estimated outputs of \textit{i)} and \textit{ii)}.   
Fig. \ref{Pos1} illustrates the localization performances of the four methods when 10\% and 20\% of all available channel data are selected for training and the rest samples are used for testing and evaluation purposes. The average distance between two adjacent training samples is smaller than $\frac{1}{8}\lambda$, i.e., the propagation channel is sufficiently sampled. As shown in Fig. \ref{Pos1}, although the truncated CIR and the raw transfer function contain the same channel information, localization performance can be improved by feeding truncated CIR into the ML pipeline. 
We postulate that it is because feeding truncated CIR into the pipeline can reduce the size of the network, which expedites the training process. Meanwhile, the signal-to-noise ratio (SNR) can be increased by truncating the tail part of CIRs, since this part contains only noise and no useful channel information. 
It can also be observed that the performance of using the instantaneous covariance matrix is better than that of the raw transfer function or truncated CIR, even though the delay information has been removed by averaging over all sub-carriers. The possible reasons are \textit{i)} the channels are LoS-dominant and the bandwidth is too small to resolve multipath components (MPCs), and \textit{ii)} the pre-processing helps the network to better exploit the angle information for localizing the UE. \textit{iii)} The network size is also decreased, which facilitates the training process. Moreover, it can be seen that the combining strategy has the best performance. This is because both delay and angle information have been utilized for training, also with the decreased network size and the increased SNR. This shows that delay information is useful for positioning, even though the bandwidth is limited. We believe that localization accuracy will be further enhanced in a rich-MPC scenario for a system with a larger bandwidth. 

To explain the accuracy improvement from another perspective, we measure the correlation coefficient $\rho$ between the positioning errors of localization methods \textit{i)} and \textit{ii)} as $\rho = \frac{\textrm{cov}(e_1,e_2)}{\textrm{var}(e_1)\hspace{1pt}\textrm{var}(e_2)}$, where $e_1$ and $e_2$ denote the Euclidean distances between $\hat{\mathbf{p}}_i^1$, $\hat{\mathbf{p}}_i^2$ and the groundtruth $\mathbf{p}_i$, respectively. The covariance between $e_1$ and $e_2$ is represented as $\textrm{cov}(e_1,e_2)$ while $\textrm{var}(e_1)$ represents the variance of $e_1$. Measurement results show that when 20\% of data is used for training, $\rho \approx 0.23$; when 10\% of data is used, $\rho \approx 0.32$. Owing to the weak correlation between $e_1$ and $e_2$, localization accuracy can be further improved by averaging $\breve{\mathbf{p}}_x^1$ and $\breve{\mathbf{p}}_x^2$. 
As shown in Fig. \ref{Pos2}, if only 2\% of channel samples are used for training, the localization accuracy of the four methods has the same ranking order as in Fig. \ref{Pos1}. However, their performances decreased significantly. We conjecture that this training density (2\%) is insufficient to represent the overall propagation channel properties since the average distance between adjacent training samples is larger than $\frac{1}{2}\lambda$, see  Fig.~\ref{Empi}. The degraded positioning accuracy can also be explained from the viewpoint of the sampling theorem, since lower-than-Nyquist sampling, i.e. too large distances between training samples, results in spectrum aliasing. 
Based on the above observations, we conclude that both data pre-processing and density of training samples play a major role in ML-based localization tasks. \vspace{-7pt}

\section{Conclusions}
In this paper, we investigate the potential of ML for indoor localization with a massive MIMO system. 
We show through a measurement campaign that pre-processing plays an important role in enhancing the ML-based localization performance. We found that in our case the two parallel networks, using instantaneous covariance matrices and truncated impulse responses as input, respectively, give two uncorrelated position estimates with similar variance, which means that they can be averaged to achieve better localization accuracy. By our approach, centimeter-level accuracy can be achieved assuming we have enough training data and sufficient channel information. To achieve this accuracy, it is not required to calibrate the whole array. If only 2\% of samples are used, median positioning accuracy is still comparable to a wavelength. In addition, it should be noted that the channel remained static and people were not present in this measurement campaign. In future works, we will investigate the influence of people and other dynamic scatters on localization accuracy, as well as other fingerprints not being so sensitive to sampling density of the training data. \vspace{-5pt} 
\label{con}
    

\setlength{\itemsep}{0em}
\renewcommand{\baselinestretch}{0.95}
\bibliographystyle{IEEEtran}
\bibliography{reference}\vspace{-2pt}

\end{document}